\documentclass{svjour3}                     % onecolumn (standard format)
\smartqed  % flush right qed marks, e.g. at end of proof
\usepackage{graphicx}
 \newcommand{\etal}{{\it et al.}}

\begin{document}

\title{CPT and Lorentz-Invariance Violation%\thanks{Grants or other notes
}


\author{Ralf Lehnert    
}


\institute{Ralf Lehnert \at
              ICN,
Universidad Nacional Aut\'onoma de M\'exico,
A.~Postal 70-543, 04510 M\'exico D.F., Mexico\\
              \email{ralf.lehnert@nucleares.unam.mx}           %  \\
          }

\date{Published: 20 August, 2009}

\maketitle

\begin{abstract}
The largest gap in our understanding of nature 
at the fundamental level
is perhaps a unified description of gravity and quantum theory. 
Although there are currently a variety of theoretical approaches to this question, 
experimental research in this field is inhibited 
by the expected Planck-scale suppression 
of quantum-gravity effects. 
However, the breakdown of spacetime symmetries
has recently been identified as a promising signal in this context: 
a number of models for underlying physics 
can accommodate minuscule Lorentz and CPT violation,
and such effects are amenable to ultrahigh-precision tests. 
This presentation will give an overview of the subject. 
Topics such as motivations, 
the SME test framework, 
mechanisms for relativity breakdown, 
and experimental tests will be reviewed. 
Emphasis is given to observations involving antimatter.
\end{abstract}

\section{Introduction}
\label{intro}

Present-day physics rests on 
two distinct theories: 
the Standard Model (SM)
describing the microscopic quantum world of elementary particles 
and General Relativity (GR) 
governing the macroscopic world dominated by gravity. 
These two theories are generally considered 
to be two low-energy aspects  
of a single, more fundamental framework 
believed to operate at the Planck scale. 
Such a fundamental framework 
must consistently unify quantum mechanics and gravity. 
Although there are numerous theoretical approaches 
to this subject,
experimental progress appears to be inhibited 
by the expected Planck-suppression 
of deviations from established physics.
Experimental quantum-gravity investigations
therefore rely largely on ultrahigh-precision searches for 
Planck-suppressed effects at attainable energies. 

Promising candidate effects 
within this context are
violations of Lorentz and CPT invariance~\cite{cpt07}.
These symmetries form cornerstones 
of both the SM and GR, 
so that any measured deviations 
necessarily imply new physics. 
Moreover, 
small Lorentz and CPT breakdown 
can be motivated by various approaches 
to physics beyond the SM and GR~\cite{lotsoftheory}.
An additional motivation for Lorentz and CPT tests 
is provided by 
the fundamental character of these symmetries: 
they should be supported as firmly as possible by
experimental evidence. 

At presently attainable energy regimes, 
such Lorentz- and CPT-breaking effects 
are described in great generality 
by the Standard-Model Extension (SME)~\cite{sme}.
The SME is a field-theory framework 
that incorporates both the usual SM and GR 
as limiting cases.
The additional Lagrangian terms of the SME 
are taken as small Lorentz- and/or CPT-violating corrections;
these corrections are constructed 
to involve all operators for Lorentz violation 
that are scalars
under coordinate changes.
This broad scope ensures 
that essentially any current or near-future experiment 
can be analyzed 
with regards to Lorentz and CPT breakdown. 
A number of studies 
have been performed within the SME~\cite{theory},
which confirm its solid theoretical foundation. 
To date, 
the SME has provided the basis 
for the identification and analysis 
of numerous experimental investigations \cite{cpt07,kr}.
For example, 
the SME leads to modifications in one-particle dispersion relations~\cite{thres}, 
which in turn could cause Cherenkov radiation in the vacuum~\cite{cher}. 
The absence of this effect at colliders 
leads to stringent limits on Lorentz violation in QED~\cite{hohen08}. 
For other limits in electrodynamics, 
see, e.g., Ref.~\cite{gamma}.

This talk is focussed on another class of tests, 
namely those involving antimatter. 
CPT symmetry implies 
that a particle and its corresponding antiparticle 
have certain identical properties, 
such as the magnitudes of mass, 
charge, gyromagnetic ratio, etc. 
This suggests 
that matter--antimatter comparisons 
can be excellent tools 
in the search for CPT violation. 
In Sec.~\ref{sec:1}, 
we review the idea behind the construction of the SME, 
and we comment on the relation 
between Lorentz and CPT symmetry.
Section~\ref{sec:2} is dedicated to phenomenology. 
In particular, 
a number of matter--antimatter comparisons 
are discussed within the context of the SME.

\section{The SME test framework}
\label{sec:1}

For the identification and analysis 
of suitable experiments, 
a test model is needed. 
The derivation of such a test model 
through a limiting process
faces various obstacles. 
One of these is the multitude of candidate underlying models 
that can accommodate Lorentz and CPT violation: 
there is presently no single realistic underlying theory, 
whose low-energy limit can serve as the test framework. 
Moreover, 
for some candidate models, 
the low-energy limit is not unique or unknown. 
For these reasons, 
the test model will be constructed by hand 
with the goal of greatest possible generality. 
This ensures the the widest  applicability 
and relative independence from the underlying physics. 

On the one hand, 
the test model should describe general breakdown of Lorentz and CPT symmetry.
On the other hand, 
this breakdown should be carefully controlled 
in the sense that
other, physically desirable properties are left unaffected. 
One of these properties is coordinate independence. 
Coordinates are a {\em mathematical tool} 
for the description of physical processes; 
they are purely a product of human thought, 
and therefore they should not acquire physical significance. 
Violating Lorentz invariance while keeping coordinate independence intact 
can be achieved by maintaining the usual Minkowski structure of spacetime 
but including preferred directions modeled by 
background vectors and tensors. 

The above low-energy description of Lorentz and CPT breaking 
with background vectors and tensors has several advantages.
First, coordinate changes are still implemented by 
the usual Lorentz transformations. 
However, 
we remind the reader that 
selecting a different coordinate system 
must be clearly distinguished from 
rotations and boosts of the experimental set-up. 
It is these rotation and boost transformations 
(i.e., the particle Lorentz tranformations), 
under which the symmetry is lost. 
The second advantage of this description is 
that it can be motivated by candidate fundamental theories:
Most approaches to underlying physics 
are based on Minkowskian manifolds 
in four or more spacetime dimensions. 
Once this structure is contained in the theory, 
it can typically not be removed 
by considering a particular low-energy solution, 
such as the vacuum. 
Indeed, 
one can think of the background vectors and tensors 
as vacuum expectation values of
Planck-mass fields. 
A third advantage is  
that a fully dynamical and microscopic description 
at presently attainable energies 
is relatively straightforward, 
as is reviewed next.

The starting point for the construction of the SME 
is essentially the entire body of established physics 
in the form of the SM Lagrangian ${\cal L}_{\rm SM}$ 
and the Einstein--Hilbert Lagrangian ${\cal L}_{\rm EH}$. 
This ensures that 
Lorentz and CPT breaking in all known physical systems 
can be accommodated. 
The next step involves adding small corrections $\delta{\cal L}_{\rm LIV}$ 
constructed by contracting the background vectors and tensors 
with ordinary SM and gravitational fields 
to form coordinate-independent scalars:
\begin{equation}
\label{smelagr}
{\cal L}_{\rm SME}={\cal L}_{\rm SM}+{\cal L}_{\rm EH}+\delta{\cal L}_{\rm LIV}\;.\end{equation}
Sample terms contained in the flat-spacetime limit of $\delta{\cal L}_{\rm LIV}$ are
\begin{equation}
\label{sampleterms}
\delta{\cal L}_{\rm LIV}\supset b_\mu\,\overline{\psi}\gamma^{\mu}\gamma_5\psi,\, 
(r_\mu\,\overline{\psi}\gamma^{\mu}\gamma_5\psi)^2,\, 
(k_F)^{\alpha\beta\gamma\delta}F_{\alpha\beta}F_{\gamma\delta},\ldots
\;.
\end{equation}
Here, $\psi$ and $F$ are a conventional spinor field
and a conventional gauge field strength, respectively. 
The nondynamical $b_\mu$, $r_\mu$, and $(k_F)^{\alpha\beta\gamma\delta}$ 
are small Lorentz-violating background vectors and tensors 
assumed to be generated by underlying physics. 
Experimental tests seek to constrain or measure these vectors and tensors.
We finally mention 
that the minimal SME (mSME) is restricted by further physical requirements, 
such as translational invariance, the usual gauge symmetries, 
and power-counting renormalizability. 
For example, the mSME does not contain the $r_\mu$ term 
present in the above expression~(\ref{sampleterms}).

In what follows, 
we focus on the mSME. 
Within the mSME, 
a subset of all Lorentz-breaking corrections 
also violates CPT symmetry. 
For instance, 
in the expression~(\ref{sampleterms}) 
the $b_\mu$ term is CPT violating, 
while the $(k_F)^{\alpha\beta\gamma\delta}$ correction 
preserves CPT. 
The questions arises, 
as to whether 
we have missed terms 
that violate CPT invariance but preserve Lorentz symmetry. 
One answer to this question is given by 
Greenberg's rigorous ``anti-CPT theorem"~\cite{anti}: 
the theorem roughly states that 
in any local, unitary, relativistic point-particle
field theory CPT violation implies Lorentz breakdown. 
{\em It follows that under these mild assumptions, 
CPT tests also probe Lorentz symmetry.}
This result offers the possibility 
for a further class of CPT-violation searches 
in addition to instantaneous matter--antimatter comparisons: 
probing for sidereal effects in matter--antimatter and other systems. 
We finally remark 
that relaxing the condition of effective unitarity, 
and thus observable probability conservation, 
can generate CPT breakdown without Lorentz violation~\cite{mav}.

\section{Experimental tests of Lorentz and CPT symmetry}
\label{sec:2}

Most theoretical mSME prediction for atomic systems 
follow similar lines of reasoning. 
The first step is the assumption of vanishing Lorentz and CPT violation 
in electrodynamics. 
This step is justified for 
most mSME coefficients in the photon sector
because astrophysical observations constrain 
their size to such a degree 
that they can be ignored for present-day atomic physics tests.
The remaining photon coefficients 
can be absorbed into other sectors of the mSME. 
The next step is the extraction of 
the modified Dirac equation
for the electron as well as the one for the proton 
(and the neutron, if needed). 
From these equations, 
the relativistic-quantum-mechanics Hamiltonian is determined. 
To do so, 
the unconventional time derivatives must be removed 
by a field redefinition~\cite{fieldredef}. 
One then proceeds 
with a generalized Foldy--Wouthuysen transformations 
that decouples the large and small spinor components.
The emerging form of the Hamiltonian
can then be used to extract the pieces for the particle and the antiparticle. 
From these, the modified nonrelativistic Pauli equation 
for the particle and the one for the antiparticle 
can be obtained~\cite{hamiltonian}. 
As a result of CPT violation, 
these two Pauli equations are inequivalent.
Lorentz- and CPT-violating corrections to atomic spectra 
can then be calculated 
employing conventional perturbation-theory methods. 
What follows is a brief description 
of various results for a number of physical systems.

{\bf The unmixed 1S--2S transition in (anti)hydrogen.} 
The experimental resolution of 
the transition involving the unmixed spin states 
is expected to be roughly one part in $10^{-18}$.
This sensitivity seems promising 
considering the likely Planck suppression for quantum-gravity effects.
However, 
the leading-order mSME analysis shows 
that there are the same shifts for free H and $\overline{\textrm{H}}$ 
in both the initial and final states 
with respect to the ordinary levels. 
As a consequence, 
this particular transition 
is less useful for detecting 
leading-order mSME effects.
Non-vanishing corrections to this transition
in the context of the mSME are generated
via relativistic effects 
that contain two further powers of 
the fine-structure constant $\alpha\simeq\frac{1}{137}$~\cite{Hbar}.
A Planck-suppressed energy shift 
would therefore exhibit a further suppression 
by a factor in excess of $10^4$. 

{\bf The spin-mixed 1S--2S transition in (anti)hydrogen.} 
For high-precision spectroscopic studies 
it is often advantageous to confine the atoms under investigation 
with magnetic fields. 
For example, 
a commonly employed set-up involves a Ioffe--Pritchard trap. 
In the present context of H and $\overline{\textrm{H}}$, 
both the 1S and 2S level are affected by 
the usual Zeeman splitting. 
An mSME calculation shows that 
the 1S--2S transition 
involving the spin-mixed states 
indeed acquires first-order corrections 
in this set-up~\cite{Hbar}.
From an experimental viewpoint, 
a potential disadvantage lies in the fact 
that this transition is also affected by the trapping magnetic field. 
It follows that 
the attainable resolution is constrained by 
the inhomogeneities in the $\vec{B}$ field.
To obtain resolutions in the vicinity of the natural line width, 
it seems likely that 
new experimental techniques must be devised.

{\bf Hyperfine Zeeman transitions within the 1S state of  (anti)hydrogen.} 
Another possibility in the context of experimental searches 
for Lorentz and CPT violation in (anti)hydrogen 
is provided by the measurement of the transition frequency 
between the Zeeman-split levels 
within the 1S state itself. 
Even for vanishing magnetic fields, 
the mSME predictions contain leading-order signals 
for such transitions~\cite{Hbar}.
Similar transitions of this kind 
(e.g., the usual Hydrogen-maser line) 
can be resolved with ultrahigh precision in experiments.
A measurement of this hyperfine Zeeman line for antihydrogen
is expected to be feasible in the near future~\cite{bj08}.

{\bf Tests involving (anti)protons in Penning traps.} 
Calculations within the mSME reveal
that not only energy levels in atoms can acquire corrections, 
but also the eigenenergies of (anti)protons 
confined in a Penning trap. 
In particular, 
one can demonstrate 
that only the $b^\mu$-type mSME parameter 
given in the expression (\ref{sampleterms})
contributes to transition-frequency differences 
between protons and antiprotons~\cite{penning}. 
More precisely, 
the anomaly transitions acquire opposite corrections 
for protons and their antiparticles.
This fact can be employed to 
extract clean experimental limits 
on the $b^\mu$ coefficient for the proton.

{\bf Searches for sidereal variations.}
In addition to instantaneous matter--antimatter comparisons 
for CPT tests, 
one can also exploit Greenberg's  ``anti CPT theorem" 
discussed in Sec.~\ref{sec:1}:
CPT breakdown comes with Lorentz violation, 
which in turn is often associated with rotation-symmetry breaking. 
It follows that 
carefully chosen measurements will then be direction dependent. 
This idea can be exploited as follows. 
A terrestrial laboratory, 
and thus the experiment, 
rotates 
as a result of the Earth spinning around its axis.
This change of orientation 
will be reflected in a roughly daily modulation 
of the experiment's observable. 
Under certain circumstances, 
higher harmonics can also occur. 
This general idea 
can be applied to a variety of physical systems. 
For example, 
modern tests  involving Hydrogen masers 
and using ingenious experimental techniques 
employ the idea of such sidereal variations~\cite{hum}.

{\bf Testing boost symmetry.} 
Lorentz invariance does not only imply isotropy 
but also symmetry under boosts. 
Paralleling the above rotation-violation searches, 
one can again exploit the motion of the Earth, 
and in particular its orbital motion. 
However, 
the Earth reverses its velocity with respect to the Sun 
about once every half year. 
This time frame could be impractical 
for experiments that
need to maintain stability throughout such periods. 
An alternative would be satellite-based Lorentz and CPT tests. 
Although there are obvious constraints 
in terms of size and weight 
as well as financial issues for such tests, 
there can also be various benefits. 
For example, 
the orbit can be selected 
with an orientation 
yielding sensitivities to different components of mSME coefficients. 
Moreover, large velocity changes 
in short amounts of time can be attained 
increasing the sensitivity to violations of boost symmetry. 
Another benefit would be the quiet environment on board a satellite, 
which clearly offers advantages for ultrahigh-precision experiments. 
In addition, 
microgravity conditions can be advantageous 
for a number of measurements. 
For example, 
in fountain-clock-type tests 
longer interrogation times, 
and thus better precision, 
can be attained for freely falling atoms.

{\bf Tests involving gravity.} 
In the experimental investigations discussed above, 
gravitational effects could be neglected 
and the flat-spacetime limit of the mSME was considered. 
However, 
tests involving gravity have recently been one focus of attention~\cite{grav,tasson,kt08}.
In particular, 
antimatter, 
such as antihydrogen, 
offers the possibility of testing Lorentz and CPT symmetry 
in the mSME's gravity sector. 
For example, 
the acceleration of antihydrogen 
in the Earth's gravitational field could be investigated. 
We also note that in gravitational contexts,
various mSME coefficients 
that are inaccessible in the flat-spacetime limit
now become measurable.
Such ideas were discussed in J.~Tasson's talk at this meeting~\cite{tasson},
and further details can be found in Ref.~\cite{kt08}.

\begin{acknowledgements}
The author wishes to thank the organizers 
for arranging this stimulating meeting 
and for the invitation to participate.
\end{acknowledgements}


\begin{thebibliography}{99}

\bibitem{cpt07}
For recent reviews of various ideas in the subject
see, e.g.,
V.A.\ Kosteleck\'y, ed.,
{\it CPT and Lorentz Symmetry I-IV},
World Scientific, Singapore, 1999-2008;
R.\ Bluhm,
Lect.\ Notes Phys.\ {\bf 702}, 191 (2006)
[hep-ph/0506054].
%%CITATION = LNPHA,702,191;%%

\bibitem{lotsoftheory}
See, e.g., V.A.\ Kosteleck\'y and S.\ Samuel,
Phys.\ Rev.\ D {\bf 39}, 683 (1989);
%%CITATION = PHRVA,D39,683;%%
V.A.\ Kosteleck\'y and R.\ Potting,
Nucl.\ Phys.\ B {\bf 359}, 545 (1991);
%%CITATION = NUPHA,B359,545;%%
S.M.\ Carroll \etal,
Phys.\ Rev.\ Lett.\ {\bf 87}, 141601 (2001);
%%CITATION = PRLTA,87,141601;%%
O.\ Bertolami \etal,
Phys.\ Rev.\ D {\bf 69}, 083513 (2004);
%%CITATION = PHRVA,D69,083513;%%
V.A.\ Kosteleck\'y, R.\ Lehnert, and M.J.\ Perry,
Phys.\ Rev.\ D {\bf 68}, 123511 (2003);
%%CITATION = PHRVA,D68,123511;%%
J.\ Alfaro, H.A.\ Morales-T\'ecotl, and L.F.\ Urrutia,
Phys.\ Rev.\ Lett.\  {\bf 84}, 2318 (2000); 
%%CITATION = PRLTA,84,2318;%%
J.D.\ Bjorken,
Phys.\ Rev.\ D {\bf 67}, 043508 (2003);
%%CITATION = PHRVA,D67,043508;%%
N.\ Arkani-Hamed \etal, 
JHEP {\bf 0701}, 036 (2007).
%%CITATION = JHEPA,0701,036;%%

\bibitem{sme}
D.\ Colladay and V.A.\ Kosteleck\'y,
Phys.\ Rev.\ D {\bf 55}, 6760 (1997);
%%CITATION = PHRVA,D55,6760;%%
Phys.\ Rev.\ D {\bf 58}, 116002 (1998); 
%%CITATION = PHRVA,D58,116002;%%
V.A.\ Kosteleck\'y and R.~Lehnert,
Phys.\ Rev.\  D {\bf 63}, 065008 (2001);
%%CITATION = PHRVA,D63,065008;%%
V.A.\ Kosteleck\'y,
Phys.\ Rev.\ D {\bf 69}, 105009 (2004).
%%CITATION = PHRVA,D69,105009;%%

\bibitem{theory} 
See, e.g., 
R.~Jackiw and V.A.~Kosteleck\'y,
Phys.\ Rev.\ Lett.\  {\bf 82}, 3572 (1999);
%%CITATION = PRLTA,82,3572;%%
V.A.~Kosteleck\'y \etal,
Phys.\ Rev.\  D {\bf 65}, 056006 (2002);
%%CITATION = PHRVA,D65,056006;%%
B.~Altschul and V.A.~Kosteleck\'y,
Phys.\ Lett.\  B {\bf 628}, 106 (2005);
%%CITATION = PHLTA,B628,106;%%
R.~Lehnert, 
Phys.\ Rev.\  D {\bf 74}, 125001 (2006);
%%CITATION = PHRVA,D74,125001;%%
arXiv:0711.4851 [hep-th];
%%CITATION = ARXIV:0711.4851;%%
A.J.~Hariton and R.~Lehnert,
Phys.\ Lett.\  A {\bf 367}, 11 (2007).
%%CITATION = PHLTA,A367,11;%%

\bibitem{kr}
V.A.~Kosteleck\'y and N.~Russell,
arXiv:0801.0287 [hep-ph].
%%CITATION = ARXIV:0801.0287;%%

\bibitem{thres}
R.~Lehnert,
Phys.\ Rev.\  D {\bf 68}, 085003 (2003).
%%CITATION = PHRVA,D68,085003;%%

\bibitem{cher}
R.~Lehnert and R.~Potting, 
Phys.\ Rev.\ Lett.\  {\bf 93}, 110402 (2004);
%%CITATION = PRLTA,93,110402;%%
Phys.\ Rev.\  D {\bf 70}, 125010 (2004).
%%CITATION = PHRVA,D70,125010;%%

\bibitem{hohen08}
M.A.~Hohensee \etal,  
Phys.\ Rev.\ Lett.\  {\bf 102}, 170402 (2009);
%%CITATION = ARXIV:0904.2031;%%
Phys.\ Rev.\  D {\bf 80}, 036010 (2009).
%%CITATION = PHRVA,D80,036010;%%


\bibitem{gamma}
V.A.~Kosteleck\'y and M.~Mewes,
Phys.\ Rev.\ Lett.\  {\bf 99}, 011601 (2007); 
%%CITATION = PRLTA,99,011601;%%
M.~Mewes,
Phys.\ Rev.\  D {\bf 78}, 096008 (2008).
%%CITATION = PHRVA,D78,096008;%%

\bibitem{anti}
O.W.~Greenberg, 
Phys.\ Rev.\ Lett.\ {\bf 89}, 231602 (2002); 
%%CITATION = PRLTA,89,231602;%%
For a somewhat more pedestrian exposition, 
see O.W.~Greenberg, 
Found.\ Phys.\  {\bf 36}, 1535 (2006).
%%CITATION = FNDPA,36,1535;%%

\bibitem{mav}
See, e.g., 
N.E.\ Mavromatos, 
these proceedings.

\bibitem{fieldredef}
R.~Lehnert,
J.\ Math.\ Phys.\  {\bf 45}, 3399 (2004).
%%CITATION = JMAPA,45,3399;%%

\bibitem{hamiltonian}
V.A.~Kosteleck\'y and C.D.~Lane,
J.\ Math.\ Phys.\  {\bf 40}, 6245 (1999).
%%CITATION = JMAPA,40,6245;%%

\bibitem{Hbar}
R.~Bluhm, V.A.~Kosteleck\'y, and N.~Russell,
Phys.\ Rev.\ Lett.\  {\bf 82}, 2254 (1999).
%%CITATION = PRLTA,82,2254;%%

\bibitem{bj08}
See, e.g., 
B.~Juh\'asz, 
these proceedings. 

\bibitem{penning}
R.~Bluhm, V.A.~Kosteleck\'y, and N.~Russell,
Phys.\ Rev.\  D {\bf 57}, 3932 (1998).
%%CITATION = PHRVA,D57,3932;%%

\bibitem{hum}
M.A.~Humphrey \etal., Phys.\ Rev.\ A {\bf 68}, 063807 (2003).

\bibitem{grav}
Q.G.~Bailey and V.A.~Kosteleck\'y,
Phys.\ Rev.\  D {\bf 74}, 045001 (2006);
%%CITATION = PHRVA,D74,045001;%%
H.~M\"uller \etal,
Phys.\ Rev.\ Lett.\  {\bf 100}, 031101 (2008);
%%CITATION = PRLTA,100,031101;%%
J.B.R.~Battat, J.F.~Chandler, and C.W.~Stubbs,
Phys.\ Rev.\ Lett.\  {\bf 99}, 241103 (2007);
%%CITATION = PRLTA,99,241103;%%
Q.G.~Bailey,
Phys.\ Rev.\  D {\bf 80}, 044004 (2009).
%%CITATION = PHRVA,D80,044004;%%

\bibitem{tasson}
J.~Tasson, 
these proceedings.

\bibitem{kt08}
V.A.~Kosteleck\'y and J.~Tasson,
Phys.\ Rev.\ Lett.\  {\bf 102}, 010402 (2009).
%%CITATION = PRLTA,102,010402;%%

\end{thebibliography}
\end{document}